\newcommand{\vdate}{April 1996}
\newcommand{\cernnr}{CERN--TH/96--107}

\newcommand{\beq}{\begin{equation}}
\newcommand{\eeq}{\end{equation}}
\newcommand{\beqn}{\begin{eqnarray}}
\newcommand{\eeqn}{\end{eqnarray}}

\newcommand{\tr}[0]{\mbox{tr}}

\newcommand{\dd}{\mbox{d}}

\newcommand{\esd}{\cally}

\newcommand{\cC}{{\esd C}}

\newcommand{\cG}{{\esd G}}
\newcommand{\cH}{{\esd H}}
\newcommand{\cL}{{\esd L}}

\newcommand{\gothn}{\gothi}

\newcommand{\gD}{{\gothn{D}}}

\newcommand{\itemsym}{{$\stackrel{\vartriangleright}{}$}}

\documentstyle[12pt,dina4p,citesort]{article}

\catcode`\@=11

 \font\tenmsx=msam10 scaled \magstep1
 \font\sevenmsx=msam8
 \font\fivemsx=msam6
 \font\tenmsy=msbm10 scaled \magstep1
 \font\sevenmsy=msbm8
 \font\fivemsy=msbm6

\newfam\msxfam
\newfam\msyfam
\textfont\msxfam=\tenmsx  \scriptfont\msxfam=\sevenmsx
  \scriptscriptfont\msxfam=\fivemsx
\textfont\msyfam=\tenmsy  \scriptfont\msyfam=\sevenmsy
  \scriptscriptfont\msyfam=\fivemsy

\def\hexnumber@#1{\ifnum#1<10 \number#1\else
 \ifnum#1=10 A\else\ifnum#1=11 B\else\ifnum#1=12 C\else
 \ifnum#1=13 D\else\ifnum#1=14 E\else\ifnum#1=15 F\fi\fi\fi\fi\fi\fi\fi}

\def\msx@{\hexnumber@\msxfam}
\def\msy@{\hexnumber@\msyfam}
\mathchardef\boxdot="2\msx@00
\mathchardef\boxplus="2\msx@01
\mathchardef\boxtimes="2\msx@02
\mathchardef\square="0\msx@03
\mathchardef\blacksquare="0\msx@04
\mathchardef\centerdot="2\msx@05
\mathchardef\lozenge="0\msx@06
\mathchardef\blacklozenge="0\msx@07
\mathchardef\circlearrowright="3\msx@08
\mathchardef\circlearrowleft="3\msx@09
\mathchardef\rightleftharpoons="3\msx@0A
\mathchardef\leftrightharpoons="3\msx@0B
\mathchardef\boxminus="2\msx@0C
\mathchardef\Vdash="3\msx@0D
\mathchardef\Vvdash="3\msx@0E
\mathchardef\vDash="3\msx@0F
\mathchardef\twoheadrightarrow="3\msx@10
\mathchardef\twoheadleftarrow="3\msx@11
\mathchardef\leftleftarrows="3\msx@12
\mathchardef\rightrightarrows="3\msx@13
\mathchardef\upuparrows="3\msx@14
\mathchardef\downdownarrows="3\msx@15
\mathchardef\upharpoonright="3\msx@16

\mathchardef\downharpoonright="3\msx@17
\mathchardef\upharpoonleft="3\msx@18
\mathchardef\downharpoonleft="3\msx@19
\mathchardef\rightarrowtail="3\msx@1A
\mathchardef\leftarrowtail="3\msx@1B
\mathchardef\leftrightarrows="3\msx@1C
\mathchardef\rightleftarrows="3\msx@1D
\mathchardef\Lsh="3\msx@1E
\mathchardef\Rsh="3\msx@1F
\mathchardef\rightsquigarrow="3\msx@20
\mathchardef\leftrightsquigarrow="3\msx@21
\mathchardef\looparrowleft="3\msx@22
\mathchardef\looparrowright="3\msx@23
\mathchardef\circeq="3\msx@24
\mathchardef\succsim="3\msx@25
\mathchardef\gtrsim="3\msx@26
\mathchardef\gtrapprox="3\msx@27
\mathchardef\multimap="3\msx@28
\mathchardef\therefore="3\msx@29
\mathchardef\because="3\msx@2A
\mathchardef\doteqdot="3\msx@2B

\mathchardef\triangleq="3\msx@2C
\mathchardef\precsim="3\msx@2D
\mathchardef\lesssim="3\msx@2E
\mathchardef\lessapprox="3\msx@2F
\mathchardef\eqslantless="3\msx@30
\mathchardef\eqslantgtr="3\msx@31
\mathchardef\curlyeqprec="3\msx@32
\mathchardef\curlyeqsucc="3\msx@33
\mathchardef\preccurlyeq="3\msx@34
\mathchardef\leqq="3\msx@35
\mathchardef\leqslant="3\msx@36
\mathchardef\lessgtr="3\msx@37
\mathchardef\backprime="0\msx@38
\mathchardef\risingdotseq="3\msx@3A
\mathchardef\fallingdotseq="3\msx@3B
\mathchardef\succcurlyeq="3\msx@3C
\mathchardef\geqq="3\msx@3D
\mathchardef\geqslant="3\msx@3E
\mathchardef\gtrless="3\msx@3F
\mathchardef\sqsubset="3\msx@40
\mathchardef\sqsupset="3\msx@41
\mathchardef\vartriangleright="3\msx@42
\mathchardef\vartriangleleft="3\msx@43
\mathchardef\trianglerighteq="3\msx@44
\mathchardef\trianglelefteq="3\msx@45
\mathchardef\bigstar="0\msx@46
\mathchardef\between="3\msx@47
\mathchardef\blacktriangledown="0\msx@48
\mathchardef\blacktriangleright="3\msx@49
\mathchardef\blacktriangleleft="3\msx@4A
\mathchardef\vartriangle="3\msx@4D
\mathchardef\blacktriangle="0\msx@4E
\mathchardef\triangledown="0\msx@4F
\mathchardef\eqcirc="3\msx@50
\mathchardef\lesseqgtr="3\msx@51
\mathchardef\gtreqless="3\msx@52
\mathchardef\lesseqqgtr="3\msx@53
\mathchardef\gtreqqless="3\msx@54
\mathchardef\Rrightarrow="3\msx@56
\mathchardef\Lleftarrow="3\msx@57
\mathchardef\veebar="2\msx@59
\mathchardef\barwedge="2\msx@5A
\mathchardef\doublebarwedge="2\msx@5B
\mathchardef\angle="0\msx@5C
\mathchardef\measuredangle="0\msx@5D
\mathchardef\sphericalangle="0\msx@5E
\mathchardef\varpropto="3\msx@5F
\mathchardef\smallsmile="3\msx@60
\mathchardef\smallfrown="3\msx@61
\mathchardef\Subset="3\msx@62
\mathchardef\Supset="3\msx@63
\mathchardef\Cup="2\msx@64

\mathchardef\Cap="2\msx@65

\mathchardef\curlywedge="2\msx@66
\mathchardef\curlyvee="2\msx@67
\mathchardef\leftthreetimes="2\msx@68
\mathchardef\rightthreetimes="2\msx@69
\mathchardef\subseteqq="3\msx@6A
\mathchardef\supseteqq="3\msx@6B
\mathchardef\bumpeq="3\msx@6C
\mathchardef\Bumpeq="3\msx@6D
\mathchardef\lll="3\msx@6E

\mathchardef\ggg="3\msx@6F

\mathchardef\circledS="0\msx@73
\mathchardef\pitchfork="3\msx@74
\mathchardef\dotplus="2\msx@75
\mathchardef\backsim="3\msx@76
\mathchardef\backsimeq="3\msx@77
\mathchardef\complement="0\msx@7B
\mathchardef\intercal="2\msx@7C
\mathchardef\circledcirc="2\msx@7D
\mathchardef\circledast="2\msx@7E
\mathchardef\circleddash="2\msx@7F
\def\ulcorner{\delimiter"4\msx@70\msx@70 }
\def\urcorner{\delimiter"5\msx@71\msx@71 }
\def\llcorner{\delimiter"4\msx@78\msx@78 }
\def\lrcorner{\delimiter"5\msx@79\msx@79 }
\def\yen{\mathhexbox\msx@55 }
\def\checkmark{\mathhexbox\msx@58 }
\def\circledR{\mathhexbox\msx@72 }
\def\maltese{\mathhexbox\msx@7A }
\mathchardef\lvertneqq="3\msy@00
\mathchardef\gvertneqq="3\msy@01
\mathchardef\nleq="3\msy@02
\mathchardef\ngeq="3\msy@03
\mathchardef\nless="3\msy@04
\mathchardef\ngtr="3\msy@05
\mathchardef\nprec="3\msy@06
\mathchardef\nsucc="3\msy@07
\mathchardef\lneqq="3\msy@08
\mathchardef\gneqq="3\msy@09
\mathchardef\nleqslant="3\msy@0A
\mathchardef\ngeqslant="3\msy@0B
\mathchardef\lneq="3\msy@0C
\mathchardef\gneq="3\msy@0D
\mathchardef\npreceq="3\msy@0E
\mathchardef\nsucceq="3\msy@0F
\mathchardef\precnsim="3\msy@10
\mathchardef\succnsim="3\msy@11
\mathchardef\lnsim="3\msy@12
\mathchardef\gnsim="3\msy@13
\mathchardef\nleqq="3\msy@14
\mathchardef\ngeqq="3\msy@15
\mathchardef\precneqq="3\msy@16
\mathchardef\succneqq="3\msy@17
\mathchardef\precnapprox="3\msy@18
\mathchardef\succnapprox="3\msy@19
\mathchardef\lnapprox="3\msy@1A
\mathchardef\gnapprox="3\msy@1B
\mathchardef\nsim="3\msy@1C
\mathchardef\napprox="3\msy@1D
\mathchardef\varsubsetneq="3\msy@20
\mathchardef\varsupsetneq="3\msy@21
\mathchardef\nsubseteqq="3\msy@22
\mathchardef\nsupseteqq="3\msy@23
\mathchardef\subsetneqq="3\msy@24
\mathchardef\supsetneqq="3\msy@25
\mathchardef\varsubsetneqq="3\msy@26
\mathchardef\varsupsetneqq="3\msy@27
\mathchardef\subsetneq="3\msy@28
\mathchardef\supsetneq="3\msy@29
\mathchardef\nsubseteq="3\msy@2A
\mathchardef\nsupseteq="3\msy@2B
\mathchardef\nparallel="3\msy@2C
\mathchardef\nmid="3\msy@2D
\mathchardef\nshortmid="3\msy@2E
\mathchardef\nshortparallel="3\msy@2F
\mathchardef\nvdash="3\msy@30
\mathchardef\nVdash="3\msy@31
\mathchardef\nvDash="3\msy@32
\mathchardef\nVDash="3\msy@33
\mathchardef\ntrianglerighteq="3\msy@34
\mathchardef\ntrianglelefteq="3\msy@35
\mathchardef\ntriangleleft="3\msy@36
\mathchardef\ntriangleright="3\msy@37
\mathchardef\nleftarrow="3\msy@38
\mathchardef\nrightarrow="3\msy@39
\mathchardef\nLeftarrow="3\msy@3A
\mathchardef\nRightarrow="3\msy@3B
\mathchardef\nLeftrightarrow="3\msy@3C
\mathchardef\nleftrightarrow="3\msy@3D
\mathchardef\divideontimes="2\msy@3E
\mathchardef\varnothing="0\msy@3F
\mathchardef\nexists="0\msy@40
\mathchardef\mho="0\msy@66
\mathchardef\thorn="0\msy@67
\mathchardef\beth="0\msy@69
\mathchardef\gimel="0\msy@6A
\mathchardef\daleth="0\msy@6B
\mathchardef\lessdot="3\msy@6C
\mathchardef\gtrdot="3\msy@6D
\mathchardef\ltimes="2\msy@6E
\mathchardef\rtimes="2\msy@6F
\mathchardef\shortmid="3\msy@70
\mathchardef\shortparallel="3\msy@71
\mathchardef\smallsetminus="2\msy@72
\mathchardef\thicksim="3\msy@73
\mathchardef\thickapprox="3\msy@74
\mathchardef\approxeq="3\msy@75
\mathchardef\succapprox="3\msy@76
\mathchardef\precapprox="3\msy@77
\mathchardef\curvearrowleft="3\msy@78
\mathchardef\curvearrowright="3\msy@79
\mathchardef\digamma="0\msy@7A
\mathchardef\varkappa="0\msy@7B
\mathchardef\hslash="0\msy@7D
\mathchardef\hbar="0\msy@7E
\mathchardef\backepsilon="3\msy@7F
\def\Bbb{\ifmmode\let\next\Bbb@\else
 \def\next{\errmessage{Use \string\Bbb\space only in math mode}}\fi\next}
\def\Bbb@#1{{\Bbb@@{#1}}}
\def\Bbb@@#1{\fam\msyfam#1}

\catcode`\@=12
\font\teneusmf=eufm10 scaled 1200
\font\seveneusmf=eufm8
\font\fiveeusmf=eufm6
\newfam\eusmffam
\textfont\eusmffam=\teneusmf
\scriptfont\eusmffam=\seveneusmf
\scriptscriptfont\eusmffam=\fiveeusmf
\def\gothi#1{{\fam\eusmffam\relax#1}}
\font\teneusm=eusm10 scaled 1200
\font\seveneusm=eusm8
\font\fiveeusm=eusm6
\newfam\eusmfam
\textfont\eusmfam=\teneusm
\scriptfont\eusmfam=\seveneusm
\scriptscriptfont\eusmfam=\fiveeusm

\font\teneusmc=cmsy10 scaled 1200
\font\seveneusmc=cmsy8
\font\fiveeusmc=cmsy6
\newfam\eusmcfam
\textfont\eusmcfam=\teneusmc
\scriptfont\eusmcfam=\seveneusmc
\scriptscriptfont\eusmcfam=\fiveeusmc
\def\cally#1{{\fam\eusmcfam\relax#1}}

\newcommand{\dgsection}[1]{\section{#1}}
\newcommand{\dgsectionstar}[1]{\section*{#1}}

\begin{document}

\begin{titlepage}

\renewcommand{\thefootnote}{\fnsymbol{footnote}}
\setcounter{footnote}{0}

\hfill
\hspace*{\fill}
\begin{minipage}[t]{4cm}
\cernnr
\newline
hep-th/9604180
\end{minipage}

\vspace{0.5cm}
\vspace{1cm}
\begin{center}

{\LARGE\bf The Quantum Gauge Principle$\;${\it\footnote[1]{{\em 
To appear in the proceedings of the Workshop on
``Time, Temporality, Now'' held at Schlo{\ss} Ringberg, Tegernsee, 
Germany, in 
February 1996.}}}
}

\vspace{0.5cm}

%
%

\vspace{1.1cm}

{\bf Dirk Graudenz}$\;${\it\footnote[3]{{\em Electronic
mail addresses: Dirk.Graudenz\char64{}cern.ch, 
graudenz\char64{}x4u.desy.de}}$\;${\it\footnote[4]{
{\em WWW URL:
http://surya11.cern.ch/users/graudenz/index.html}}} \\
\vspace{0.1cm}
Theoretical Physics Division, CERN\\
CH--1211 Geneva 23\\
}

\end{center}

\vspace{1.5cm}
\begin{abstract}
We consider the evolution of quantum fields on a classical background 
space-time, formulated in the language of differential geometry. Time evolution
along the worldlines of observers is described by parallel transport operators 
in an infinite-dimensional vector bundle over the space-time manifold.
The time evolution equation and the dynamical equations for the matter fields
are invariant under an arbitrary local change of frames along the restriction 
of the bundle to the worldline of an observer, thus implementing a ``quantum 
gauge principle''. We derive dynamical equations for the connection and a 
complex scalar quantum field based on a gauge field action. In the limit of 
vanishing curvature of the vector bundle, we recover the standard equation
of motion of a scalar field in a curved background space-time.
\end{abstract}

\renewcommand{\thefootnote}{\arabic{footnote}}
\setcounter{footnote}{0}

\vfill
\noindent
\begin{minipage}[t]{5cm}
\cernnr\\
\vdate
\end{minipage}
\vspace{1cm}
\end{titlepage}


\newpage

\dgsection{Introduction}
\label{intro}
\noindent
The concept of time in quantum field theory is derived from the
structure of the underlying space-time manifold. For both flat and 
curved space-times, in the 
Heisenberg picture, the state vector of a quantum system is 
constant\footnote{We do not consider the problem of measurements.} and
identical
for all observers, and the quantum fields fulfil dynamical equations
derived by means of a quantization procedure of classical equations
of motion based on a classical action functional \cite{1}. 
This framework is unsatisfactory for two principal reasons:
\begin{itemize}
\item[(i)] The state vector is an object that describes the knowledge
of an observer about a physical system. 
The observer, in an idealized case, moves along a worldline $\cC$.
The state vector should thus be tied to $\cC$, and its time evolution
should not be related to some globally defined time, but to the
{\it eigenzeit} of the observer.
\item[(ii)] The observer is free to choose the basis vectors in Hilbert space.
In quantum mechanics, a change of basis vectors 
amounts to a change of the ``picture'', e.g.\ 
from the Heisenberg to the Schr\"odinger picture. In quantum field 
theory, in particular in perturbation theory, the preferred picture is 
the interaction picture. The change of pictures is performed by means
of unitary operators. If, as proposed in (i), the state vector is 
tied to a worldline, a change of the basis in the Hilbert space
should be allowed to be observer-dependent and arbitrary at any point 
of the worldline. 
\end{itemize}
These two requirements amount to what could be called a 
``local quantum relativity
principle'': physics is independent of the Hilbert space 
basis, and the basis may be 
chosen locally in an arbitrary way. 
Since a change of the basis is mediated by a unitary transformation, 
the local quantum relativity principle is equivalent to a local
$U(\cH)$ symmetry, $U(\cH)$ being the group of unitary operators
on the Hilbert space $\cH$.

\medskip
\noindent
In Ref.~\cite{2}, a formulation of the time evolution of quantum systems
in the language of differential geometry has been given.
State vectors are elements of an infinite-dimensional vector bundle, 
and time evolution is given by the parallel transport operator 
related to a connection in the bundle.
In order to keep the paper self-contained, this framework is briefly
reviewed in Section~\ref{diffgeom}.
The goal here
is to propose a dynamical principle that 
yields the connection in the bundle and the field operators at any 
space-time point. The basic idea is that the geometrical formulation
permits the introduction of a local gauge theory, the connection being the
``gauge field'', and the quantum field operators being linear-operator-valued 
``matter fields''. Such a local gauge theory will be defined in 
Section~\ref{localgauge} by means of an action functional.
We wish to note that the theory does not have to be quantized, because
the dynamical variables appearing in the action
are already the components of the quantum operators 
in an arbitrary frame. It is, however, not yet clear whether in this 
way canonical commutation relations hold in general.

\medskip
\noindent
The theory distinguishes conceptually between the evolution of the 
state vector, given by the connection in the bundle (for short, 
called the ``quantum connection'' in the following), and the dynamical 
equations of the quantum fields and the quantum connection, derived from 
the action principle. The time evolution of the state vector, bound to the 
worldline of a 
specific observer, and the space-time dependence of the quantum fields 
are in principle 
independent, although the generator of the state-vector time evolution
is a field that will appear in the equation of motion of the quantum 
fields.
Since the action and the time evolution equation 
are formulated in a gauge-covariant way, the local quantum relativity principle
is fulfilled.

\medskip
\noindent
The paper closes with a discussion. Some technical remarks are relegated
to an appendix.

\dgsection{The Geometrical Framework}
\label{diffgeom}
\noindent
In this section, we briefly review the geometric formulation 
of quantum theory. For more details, we refer the reader to 
Ref.~\cite{2}. The appendix contains a short collection of 
useful relations and other technical details.

\medskip
\noindent
The basic ingredients
of the theory are\footnote{
In the following, we do not distinguish in the notation 
between the global, coordinate-independent quantities and their 
basis-dependent components in, for example, a local trivialization
of the vector bundle.
What is meant will be clear from the context.
Moreover, we do not attempt to fulfil any standards of mathematical
rigour; the focus is on the conceptual development.  
For example, we do not discuss the problems coming from 
the regularization of operator products.
We also sometimes drop technical details. It is, for instance,
assumed that local quantities are patched together by a partition 
of unity, whenever this is required.
}:
\begin{itemize}
\item[(a)] A space-time manifold $M$ with metric $g_{\mu\nu}$.
\item[(b)] A Hermitian vector bundle 
$\pi:H\rightarrow M$ over M, with the fibres $H_x=\pi^{-1}(x)$ 
isomorphic to a Hilbert space $\cH$. The structure group
of the bundle
is assumed to be the unitary group $U(\cH)$ of the Hilbert space,
and Hermitian conjugation in a local trivialization is denoted by~`$*$'.
The metric of the bundle is
denoted by $G$, and Hermitian conjugation with respect to $G$ is
denoted by `$\dagger$'.

\item[(c)] The bundle $H\rightarrow M$ is equipped with a connection $D$. 
In a local 
trivialization, the connection coefficients are denoted by $K$;
they are 
anti-Hermitian operators 
\beq
\label{antih}
K=-K^\ast.
\eeq
The covariant derivative
of a section $\psi$ of the bundle is defined by 
\beq
\label{covder}
D\psi=\dd{}\psi-K\psi.
\eeq 
The curvature $F$ corresponding to $D$ is $F=D^2$.
The covariant derivative of a field $A$ of linear maps of the fibres
of $H\rightarrow M$ is given by 
\beq
\label{DA}
DA=\dd A-KA+AK,
\eeq 
and similarly the covariant 
derivative of the metric reads
\beq
\label{DG}
DG=\dd G-KG+GK.
\eeq 
It is assumed
that the connection and the metric are compatible, i.e.\
\beq
\label{DG0}
DG=0.
\eeq
\end{itemize}

\medskip
\noindent
The physical interpretation of these mathematical objects and some assumptions
being made are:
\begin{itemize}
\item[(a$^\prime$)] 
The underlying space-time manifold $M$ together with its metric
is assumed to be fixed and given by some other theory. 
It would certainly be desirable that the 
dynamical laws governing the evolution of the metric field, i.e.\ the
Einstein equations, could be incorporated in the formalism developed here.
This could be achieved, for example, by adding the Einstein--Hilbert action
$S_{EH}$
to the quantum action $S_Q$ to be defined later\footnote{
Should a genuine quantum theory of gravity
be possible, then the
theory developed here could certainly no longer be applied,
because in the differential
geometric formulation we make use of the fact that the manifolds
and bundles under 
consideration are smooth. What would be required 
in this case would be a geometry of
space-time compatible with quantum gravity.

It is not clear {\it a priori} whether a quantum theory of gravity
can be formulated by quantizing some classical action. 
Conceptually, space-time is the set of all possible
events, and the metric, up to a conformal factor, merely encodes the
causal structure. Epistemologically, these notions are much more fundamental
and deeper
than gauge and matter fields.
It is possible that gravity should not be quantized at all.
}.
\item[(b$^\prime$)] 
For an observer $B$ at $x\in M$ we assume that the state vector $\psi$
that $B$ uses as a description of the world is an element of 
$H_x$. Observables, such as fields $\Phi(y)$, are assumed to be 
sections of the bundle $\pi_{\cL H}:\cL H\rightarrow M$, whose fibres
consist of linear operators acting on the fibres of $H\rightarrow M$.
We use the bracket notation for the inner product given by $G$; 
for two vectors $\xi$, $\eta\in H_x$ and a linear operator
$A\in \cL H_x$, we write 
$G(\xi,A\eta)=\langle\xi|A|\eta\rangle$.
The quantity\footnote{We do not include the space-time
point $x$ in the notation for the metric $G$.} 
$\langle\psi|\Phi(x)|\psi\rangle$, for $\psi\in H_x$, is assumed
to be the expectation value of the field $\Phi$ at $x$, i.e.\ the
prediction for the average of a measurement by means of a local
measurement device carried by $B$. 
Predictions by an observer $B$ at $x$ for a 
measurement of $\Phi$ at $y$ can be made by a parallel transport 
of the state
vector 
along a path joining $x$ and $y$ 
(see (c)), and taking the expectation value at $y$. Requiring 
path independence of the expectation value, i.e.\ consistency of
predictions, leads to a condition
\beq
\label{ccond}
\left[U_\alpha,\Phi(x)\right]\,\psi=0
\eeq
for state vectors $\psi$ of the physical subspace of $\cH_x$, 
closed loops~$\alpha$
attached to $x$ and observables $\Phi(x)$.
There is thus a symmetry group (the group of $U_\alpha$ fulfilling 
Eq.~(\ref{ccond})) related to the holonomy group of the bundle;
for a discussion see Ref.~\cite{2}\footnote{
Invariant operators $\breve{A}\in \cL H_x$ fulfilling 
$[U_\alpha,\breve{A}]=0$
for all closed curves $\alpha$ can be constructed from an arbitrary
operator $A\in \cL H_x$ by means of a ``path integral'' 
$\breve{A}=\int \gD\beta\,U_\beta\, A \,U_\beta^{-1}$ 
over all closed curves $\beta$
originating in $x$, if a left-invariant
($\int \gD\beta\,f(\alpha\circ\beta)=\int \gD\beta\,f(\beta))$ and
normalized ($\int \gD\beta=1$) measure $\gD\beta$ exists.
}.
\item[(c$^\prime$)] 
The quantum connection $D$ can be integrated along curves $\cC$
joining $x$ and $y$
to give parallel
transport operators $U_{\cC}$ mapping the fibre $H_x$ onto the fibre $H_y$.
The quantum connection $D$ is assumed to govern the evolution of the
state vector $\psi$ in the direction of a tangent vector $v$ 
by means of the equation $D\psi(v)=0$. 
For an observer $B$ moving along a worldline $\cC(\tau)$, parametrized by 
$B$'s eigentime $\tau$, the evolution of the state vector $\psi(\tau)$
is thus 
\beq
\label{seq}
\partial_\tau\psi(\tau)=K_\mu(\cC(\tau))\dot{\cC}^\mu(\tau)\,\psi(\tau).
\eeq
This equation is nothing but a Schr\"{o}dinger equation\footnote{
It is possible to include, for example, a one-form $P(\Phi,D\Phi,F)$,
polynomial in the fields $\Phi$, the derivatives $D\Phi$ and the
curvature $F$, in the evolution equation, such that
$(D-P)\Psi(v)=0$. This would correspond to an
additional interaction term in the Schr\"odinger equation.}
for a path-dependent
Hamilton operator\footnote{To simplify the notation and suppress factors
of the imaginary unit, we require the operator $K$ to be anti-Hermitian,
see Eq.~(\ref{antih}).}
or ``quantum gauge field'' $K_\mu$.
The assumption of $D$ and $G$ being compatible means that the transition 
amplitude $\langle\psi(\tau)|\chi(\tau)\rangle$ of two states $\psi$ and
$\chi$ is invariant under time evolution.
\end{itemize}

\medskip
\noindent
We now have to discuss the question 
of where the quantum connection $D$ and the 
dynamical equation for the quantum matter fields $\Phi$ come from.
It is desirable to have a common principle for these two objects.
We note that the matter fields are related to the vector 
bundle itself, whereas the
connection is naturally related to the principal bundle.
We therefore need a means to connect objects related to two different
bundles.
The following possibilities suggest themselves:
\begin{itemize}
\item[(A)] If there is a preferred trivialization of the bundle, i.e.\ 
a canonical coordinate system, then in this particular system the 
quantum 
connection coefficients $K$ can be defined as a function of the 
matter fields $\Phi$.
An example of this is the translation of 
the standard formulation of quantum field theory in the Heisenberg picture
in Minkowski space to 
the geometric formulation. The bundle $H\rightarrow M$ 
is nothing 
but the direct product $M\times \cH$ of Minkowski space $M$ and the
Hilbert space $\cH$. There is a canonical trivialization
of the bundle owing to the direct product structure. 
The quantum connection coefficients $K$ are set to zero
in this trivialization; 
$D$ is thus simply the total differential. 
Consequently, the state vector is
constant and the same for all observers. The metric $G$ is inherited from 
the metric of $\cH$. The dynamical law for the fields $\Phi(x)$ is the
Heisenberg equation of motion 
\beq
\label{heisenberg}
\partial_\mu\Phi(x)=i\,[P_\mu,\Phi(x)],
\eeq 
where
the $P_\mu$ are the energy and momentum operators of the 
theory, being functions
of the fields $\Phi$.
The crucial step in this construction is the assumption of a trivial 
bundle $M\times\cH$, because with this
a preferred trivialization of the bundle
comes for free.
\item[(B)] A variant of (A) is to single out a specific coordinate 
system by some physical principle,
the prototype being the definition of inertial frames
and the application of the equivalence principle
in general relativity. Unfortunately, the application of an equivalence
principle based on inertial frames is not possible in our case, because
this would only fix a frame in the tangent bundle of $M$, but not 
in the bundle $H\rightarrow M$.
\item[(C)] Finally, there is the possibility to postulate a dynamical law.
This is well suited to the problem at hand, because the connection is 
essentially a differential operator on the vector bundle. This allows 
us to define covariant differential equations possibly derived from
an action principle.
\end{itemize}
In this paper, we follow (C) by defining a gauge field 
action for the special case of a 
complex scalar field $\Phi$.
This and the derivation of the dynamical equations is done in the next section.
 
\dgsection{The Quantum Action and the Dynamical Equations}
\label{localgauge}
\noindent
The action employed to derive dynamical equations for the 
quantum connection~$D$
and a complex scalar quantum field~$\Phi$ is
\beq
\label{SQ}
S_Q=S_K+S_G+S_k+S_m,
\eeq
where 
\beq
\label{SH}
S_K=\int\dd x\,\sqrt{\sigma g}\,
\frac{\alpha}{2}\,\tr\left(F_{\mu\nu}F^{\mu\nu}\right)
\eeq
is the action for the quantum connection coefficients,
\beq
\label{SG}
S_G=\int\dd x\,\sqrt{\sigma g}\,
\tr\left(\lambda^\mu D_\mu G\right)
\eeq
is the action implementing the constraint $DG=0$ by means of a field
of linear-operator-valued Lagrange multipliers $\lambda^\mu$,
\beq
\label{Sk}
S_k=\int\dd x\,\sqrt{\sigma g}\,
\tr\left(\left(D_\mu\Phi\right)^\dagger D^\mu\Phi\right)
\eeq
is the action for the kinetic part of the field $\Phi$, and
\beq
\label{Sm}
S_m=\int\dd x\,\sqrt{\sigma g}\,
\gamma\,\tr\left(\Phi^\dagger \Phi\right)
\eeq
is an action reminiscent of a mass term for $\Phi$.
Here $\sigma$ is the sign of the determinant 
\beq
\label{detg}
g=\mbox{det}\left(g_{\mu\nu}\right)
\eeq
of the
space-time metric, $\alpha$ and $\gamma$ are ``coupling constants'' to 
be discussed later, and $F_{\mu\nu}$ is the curvature tensor associated with
the quantum connection~$D$, defined by
\beq
\label{curv}
F_{\mu\nu}=-\partial_\mu K_\nu+\partial_\nu K_\mu+[K_\mu,K_\nu].
\eeq
The unusual signs in the first two terms stem from the fact that 
$F=D^2$, where $D=\dd-K$
instead of $D=\dd+K$, as is usually assumed.
The trace `$\tr$' is the trace operation of linear operators
in a local trivialization of 
the vector bundle. 

\medskip
\noindent
It can easily be checked that the action $S_Q$ is invariant under 
a change of basis in the vector bundle. Owing to the Lagrange 
multipliers $\lambda^\rho$, the variables $K_\rho$, $G$, $\Phi$, $\Phi^*$ and
$\lambda^\rho$ are independent. The action principle $\delta S_Q=0$ then 
leads to the equations of motion by varying the fields\footnote{The variations
$i\,\delta K_\rho$, $\delta G$, $\delta\Phi$, $\delta\Phi^*$ and
$\delta\lambda^\rho$ run through all infinitesimal Hermitian operators 
$\delta R$, 
so that the condition $\tr\left(A\,\delta R\right)=0$ for all $\delta R$
leads to $A=0$.}.
In order to achieve a compact notation, we introduce the 
covariant
derivative $\hat{D}_\mu$
for a vector $t^\mu$ and for an antisymmetric 
tensor $t^{\mu\nu}$ by
\beq
\label{hatd}
\hat{D}_\mu\, t^\mu=\frac{1}{\sqrt{\sigma g}}
D_\mu\left(\sqrt{\sigma g}\,t^\mu\right)
\eeq 
and
\beq
\label{hatd2}
\hat{D}_\mu\, t^{\mu\nu}=\frac{1}{\sqrt{\sigma g}}
D_\mu\left(\sqrt{\sigma g}\,t^{\mu\nu}\right),
\eeq 
respectively. It can be shown that for vanishing 
torsion these expressions are 
covariant divergences, and transform as a scalar and as a vector, 
respectively.

\begin{itemize}
\item[($\alpha$)]
The variation of the quantum connection 
coefficients~$K_\rho$ leads to\footnote{
This is an equation for the covariant derivative $D\,{\ast F}$ of the 
dual curvature tensor $\ast F$. 
The explicit form of $D\,{\ast F}$ is given by 
Eq.~(\ref{DA}).
There is, of course, also the Bianchi identity $DF=0$.
}
\beq
\label{eqofmotion1}
\frac{\alpha}{2}\,\hat{D}_\mu F^{\mu\rho}+\left[\lambda^\rho,G\right]
-G\,\left[\Phi^\dagger,D^\rho\Phi\right]\,G^{-1}
-\left[\Phi,\left(D^\rho\Phi\right)^\dagger\right]=0.
\eeq
In a classical gauge theory, if the vector bundle is finite-dimensional,
this is the equation of motion for a gauge
field coupled to a matrix-valued 
complex scalar field in the fundamental representation. 
In our case, the gauge field is
related to the quantum connection~$D$. It should be noted that this
equation is different from the one obtained when quantizing, for example,
a classical 
$SU(n)$ gauge field in conjunction with a matrix-valued matter field in the 
fundamental representation. In this case, for the gauge field, 
we would have operators $A_{\mu a}$, where $a$ is a colour index in the
adjoint representation. The matter field ${\Phi^b}_c$
would come with colour indices~$b$ and~$c$ in the fundamental representation.
In Eq.~(\ref{eqofmotion1}), 
the operators do not carry an explicit colour index, 
rather the ``colour indices'' are the indices of the infinite-dimensional
matrices, if the equations were written out in a specific Hilbert space basis.

\item[($\beta$)]
The variation of the metric $G$ results in
\beq
\label{eqofmotion2}
\left(\hat{D}_\mu\lambda^\mu\right)G
+\left[\left(D_\mu\Phi\right)^\dagger,D^\mu\Phi\right]
+\gamma\,\left[\Phi^\dagger,\Phi\right]=0.
\eeq
The solution of this equation yields the Lagrange multiplier $\lambda$, 
eventually to be inserted into the other equations.

\item[($\gamma$)]
The variation of the complex scalar field $\Phi$ leads to
\beq
\label{eqofmotion3}
\hat{D}_\mu D^\mu\Phi-\gamma\,\Phi=0
\eeq
and
\beq
\label{eqofmotion4}
\hat{D}_\mu \left(D^\mu\Phi\right)^\dagger-\gamma\,\Phi^\dagger=0.
\eeq
To discuss these equations, let us set the quantum 
gauge field in the covariant derivative to 
zero\footnote{This corresponds to $F_{\mu\nu}=0$.}.
This can be achieved by the limit 
$\alpha\rightarrow\infty$, for the following reason.
Defining $\alpha=1/a^2$ and $\tilde{K}_\mu=K_\mu/a$ allows 
the coefficient $\alpha$ 
to be absorbed 
into the curvature tensor $\tilde{F}_{\mu\nu}$
of $\tilde{K}_\mu$, where
the commutator term in $\tilde{F}_{\mu\nu}$ receives a factor of $a$.
The covariant derivative is $D=\dd-a\tilde{K}$. Setting $a=0$ leads
to the desired result.
Equation~(\ref{eqofmotion3}) then reduces to 
\beq
\label{eqofmotion3p}
(\Box-\gamma)\Phi=0,
\eeq
with
\beq
\label{KG}
\Box\,\Phi=\frac{1}{\sqrt{\sigma g}}\,
\partial_\mu\left(\sqrt{\sigma g}\,g^{\mu\nu}
\partial_\nu\Phi\right)
\eeq
the wave operator on the space-time manifold~$M$.
Defining $\gamma=-m^2$, Eq.~(\ref{eqofmotion3p}) is the Klein--Gordon equation
for a scalar quantum field of mass $m$ in a curved background space-time
\cite{1}. 
Moreover, for $a=0$ the state vector is constant along the worldline
of the observer.
We are thus able to recover standard quantum field theory 
in curved background space-times in a certain limit.

\item[($\delta$)]
Finally, the
variation of the Lagrange multipliers $\lambda^\rho$ yields the constraint 
that the metric in the vector bundle be consistent with the quantum connection:
\beq
\label{eqofmotion5}
D_\rho G=0.
\eeq
\end{itemize}

\medskip
\noindent
As can easily be seen, the equations of motion are all explicitly
gauge covariant.

\dgsection{Discussion}
\label{discussion}
\noindent
In this paper, we have proposed dynamical equations for the geometrical 
formulation of quantum field theory as defined in Ref.~\cite{2}.
A classical gauge field action for a complex scalar field in an 
infinite-dimensional vector bundle\footnote{To be more precise, for a complex
scalar field in the bundle of linear operators acting on the fibres of a
vector bundle.} gives rise to gauge covariant equations of motion for
the quantum connection and for the scalar field.
A gauge transformation can be interpreted as a change of frame in the 
vector bundle, and thus as a space-time-dependent change of the ``picture''.

\medskip
\noindent
We wish to point out a similarity 
of the present theory to the quantum mechanics of a single 
particle coupled to an electromagnetic field. 
There, the requirement that the phase of the wave function have
no physical meaning motivates the introduction of an Abelian gauge field, 
which can then be interpreted as the electromagnetic gauge potential.
The Schr\"{o}dinger wave function is in general
a section of a complex line bundle. Unobservability of the 
phase can be rephrased as the independence of physics
of the particular choice of basis in the line bundle, admitting arbitrary
passive space-time-dependent $U(1)$ transformations.
In our case, the situation is slightly
different. We are not concerned with the quantum mechanics of a single
particle, but with quantum field theory, where, in the geometrical 
formulation, the space-time dependence
relates to the full state vector and not only to the amplitude at a specific
space-time point. The quantum relativity principle states that physics 
be independent of the choice of basis in the Hilbert space, for all 
possible observers. Instead of the independence of physics 
of the phase of the 
Schr\"odinger wave 
function, we
require that physics 
be invariant under arbitrary local $U(\cH)$ transformations.
Since the Abelian gauge potential in the quantum mechanics case actually
relates to an empirically observable field, it is tempting to speculate
whether the quantum gauge connection has some counterpart in physical
reality as well. The fact that a dynamical formulation 
involving the
quantum connection coefficients, as done in this
paper, is possible, and consequently a resulting 
set of coupled equations of motion of the 
quantum connection coefficients and the matter fields can be derived,
suggests that
quanta of the matter fields can, by quantum fluctuations, be 
transformed into (hypothetical) quanta of the quantum connection.

\medskip
\noindent
The gauge theory structure of the geometrical formulation 
naturally leads to some additional questions:
\begin{itemize}
\item[\itemsym] 
Are there conserved Noether currents, 
and if so, how should these be interpreted?
\item[\itemsym] 
In the present context, ``gauge fixing'' means
the choice of a particular local trivialization of the vector bundle.
Locally, the quantum gauge field $K_\mu$ can be transformed to zero
if and only if the curvature $F_{\mu\nu}$ vanishes.
In general, this is not the case. However, for an observer $B$, it is
always possible to choose a specific frame such that $K_\mu=0$ on the 
restriction of the bundle to the worldline. This corresponds to a Heisenberg
picture for $B$, since the state vector will be constant.
In general, an additional condition $\partial_\mu K_\nu=0$ along the worldline
cannot be achieved. Locally, therefore, the quantum gauge field cannot 
be transformed away, and this raises the question of its
physical significance.
\end{itemize}

\medskip
\noindent
Another set of questions relates to quantum field theory aspects: 
\begin{itemize}
\item[\itemsym]
What are
suitable initial conditions for the equations of motion?
\item[\itemsym]
How should locality of fields be defined in the geometrical formulation?
\item[\itemsym]
Is it possible to define a vacuum state? Is the vacuum state dependent on the
state of motion of the observer (for example, does the Unruh effect lead 
to a different vacuum state for an accelerated observer)?
\item[\itemsym]
Is it possible mathematically to make sense out of the theory;
for example, can a perturbative expansion in the coupling
constant~$a$ of the quantum connection be derived?
\item[\itemsym]
Is there a way to introduce self-interacting scalar fields, fermions and
``ordinary'' gauge fields in the quantum action?
\end{itemize}

\medskip
\noindent
We close the discussion with a remark concerning a ``global Hamiltonian''.
In order to recover Heisenberg type
equations of motion for the quantum gauge field 
and for the matter fields, 
we need a ``global Hilbert space''. The bundle 
itself can be considered to be 
such an object. Elements of this space are bundle sections
$\xi$, and a global metric can be defined by
\beq
\label{globG}
\cG\left(\xi,\eta\right)=\int\dd x \,\sqrt{\sigma g}\,
G\left(\xi(x),\eta(x)\right).
\eeq
An interesting problem would be to find a global operator $R$,
mapping sections of the vector bundle into vector-valued one-forms,
such that an equation of the type
\beq
\label{heisenbergp}
D\varphi=i\,[R,\varphi]
\eeq
hold for all fields $\varphi$ of the theory, including the quantum gauge field,
under the assumption of suitable commutation relations.

\dgsectionstar{Acknowledgements}
\noindent
I am grateful to H.~Atmanspacher and E.~Ruhnau for 
inviting me to participate in a very interesting workshop.

\dgsectionstar{Appendix}
\label{appa}
\noindent
In this appendix, we collect some technical details related to 
differential geometry. General references are \cite{3} and
\cite{4}. 

\medskip
\noindent
In a local trivialization of the vector bundle, the Hermitian conjugate 
$A^\dagger$ of a linear operator $A$ with respect to the metric $G$ is
given by
\beq
\label{hcg}
A^\dagger=G^{-1}\,A^\ast\,G.
\eeq

\medskip
\noindent
A change of frame 
\beq
\label{vb}
\xi^\prime=S\,\xi
\eeq
in the vector bundle, $S$ being a 
unitary operator, leads to a transformation of linear operators
of the form 
\beq
\label{linop}
A^\prime=S\,A\,S^{-1}
\eeq
and to a transformed connection $D^\prime=\dd-K^\prime$ with
\beq
\label{hop}
K^\prime=S\,K\,S^{-1}-\dd S\,S^{-1}.
\eeq
The curvature $F_{\mu\nu}$ transforms as
\beq
\label{curvvar}
F_{\mu\nu}^\prime=S\,F_{\mu\nu}\,S^{-1}.
\eeq

\medskip
\noindent
A variation of 
$\tr\left(A^\dagger B\right)$ with respect to the metric $G$,
keeping $A^\ast$ and $B$ fixed,  leads to
\beq
\label{trvar}
\delta_G\,\tr\left(A^\dagger B\right)
=\tr\left(-\left[A^\dagger,B\right]\,G^{-1}\delta G\right).
\eeq
This expression is useful for the derivation of the equations of motion.


\newcommand{\scs}{\rm}
\newcommand{\bibitema}[1]{\bibitem[#1]{#1}}
\newcommand{\bibbeginlong}{\begin{thebibliography}{XXX\,1988\,\,}}
\newcommand{\bibbeginshort}{\begin{thebibliography}{XX}}

\bibbeginshort 

\bibitema{1}
   {\scs N.D.~Birrell and P.C.W.~Davies,}
   {\it Quantum Field Theory in Curved Space}
   (Cambridge University Press, Cambridge, 1982).

\bibitema{2}
   {\scs D.~Graudenz,}
   {\it On the Space-Time Geometry of Quantum Systems},
   {preprint CERN--TH.7516/94 (November 1994)}.

\bibitema{3}
   {\scs S.~Kobayashi and K.~Nomizu,}
   {\em Foundations of Differential Geometry}
   (John Wiley \& Sons, New York, 1963).

\bibitema{4}
   {\scs R.O.~Wells,}
   {\em Differential Analysis on Complex Manifolds}
   (Springer Verlag, New York, 1980).
 
\end{thebibliography}

\end{document}